\documentclass[12pt,a4paper]{article}
\usepackage{graphicx,amssymb,bm,latexsym,amsmath,braket,wrapfig,cite,color}
\usepackage{float}
\makeatletter
\@addtoreset{equation}{section}

\makeatother

\newcommand{\bea}{\begin{eqnarray}}
\newcommand{\eea}{\end{eqnarray}}

\newcommand{\pa}{\partial}
\newcommand{\til}{\widetilde}
\newcommand{\nn}{\nonumber\\}
\newcommand{\p}[1]{(\ref{#1})}

\newcommand{\tr}{\mathrm{Tr}}

\begin{document}
\begin{center}
{\fontsize{18}{0pt}\selectfont\bf
Renormalization group of the gravity coupled with the scalar theory and its effective normalization} \\

\end{center}

\medskip
\renewcommand{\thefootnote}{\fnsymbol{footnote}}

\begin{center}
Daiki Yamaguchi
\footnote{e-mail: daichanqg@gmail.com} \\

\medskip

\emph{${}^1$Department of Physics, Kindai University, Higashi-Osaka, Osaka 577-8502, Japan
\\
\medskip
${}^2$Research Institute for Science and Technology, \\
Kindai University, Higashi-Osaka, Osaka 577-8502, Japan
\\
\medskip
${}^3$Related with Research Institution of Science and its Departments, \\
Kyosan University, Kamikaga-Kyoto, Kyoto, Japan}

\end{center}
\medskip

\begin{center}
{\bf Abstract}
\end{center}
The normalization of the quantum corrected action is resolving the equation divergent dependence of the cutoff towards the system apparent result in quantum gravity. Here we consider the normalization to Einstein R twice scalar action with the cutoff runs from apparent infrared momentu to the ultraviolet momentum. These gravitational actions, Einstein R twice and Einstein R twice with the scalar theory, are come to the ensured apparent system recommendations in classical quantum gravity.

\renewcommand{\thefootnote}{\arabic{footnote}}

\section{Introduction}
\label{sec:1}
The quantum correction renormalization is the problem in modern physics. The 1 loop quantum correction and anomalous dimension dependent is researched in many works such as \cite{Percacci} these have a lot of methods for thinking about the renormalization with the functional renormalizationg group equation. From \cite{Percacci}, the gravitational renormalization is used with the local heat kernel expanssion and its coefficients are list. The quantum gravity meaned as the field theory with gravitational perturbative expanssion is renormalized with the heat kernel expanssion of the renormalization group equation, and so, the equation divergence cutoff scale $k^2$ is absorbed into the running coupling as $G(k\to\Lambda)$. In moder research \cite{DEP} is the representative work for renomalized the quantum gravity with matters numbers towards Newton coupling goes to asymptotic safe from the classic $G_0$. Fixed point behaives $0<g_*<1$.

In this study, we consider the nonlocal heat kernel expanssion of the functional renormalization group equation and its expanded result of the quantum gravity which is the gravity coupled with the scalar theory \cite{CPRT16}, and then, we give the normalize partition function lead the effective action towards the gravitational effective action with background metric and scalar field. Since the gravity and scalar actions are perturbativelly expanded to fields quadratic forms, These are summerized in to the field $\delta\Phi$. The Hessians these are gravity coupled scalar and ghost are derived with the decided backgroud metric and scalar. The functional renormalization group equation with non local heat kernel expanssion lead to the effective action with verifications of the heat kernel coefficients and structure functions. Therefore, the gravitational effective actions these are scalar coupled Einstein $R^2$ action and Einstein $R^2$ action with no scalar matter.

The normalized actions \p{gsef} and \p{gref} are the gravitational effective action the constants are positivelly decided as to be sure the appropriate actions with the proofs of the gravity. From these reasons, we are ensure that the normalized gravitational actions are having possibilities to find fits of the gravity datas. Einstein $R^2$ scalar action or Einstein $R^2$ action are matched to the needed sophisticated systems. The cosmologies and gravitation proplems are anticipated to be resolved with these Einstein $R^2$ types actions as the classical quantum gravity.

This paper is list contents as followings. In section 2, we consider the non local heat kernel expanssion of the functional renormalization group equation with the scalar theory on the Einstein manifold. The generally also formally master equation and structure functions are verified in this section. In section 3, we consider the scalar gravity expanded to the fields theory as dependent of the laplace operator and that's ghost operator. The non local heat kernel expanssion with heat kernel coefficients and structure functions are to the renormalization group equation with gravity coupled with the scalar. The effective action is got by the cutoff scale integration these are $\mu(>\sqrt{z}/2)\to\Lambda$ \cite{CPRT16}. Over the reference, we could give the partition function as $Z$. The given effective action is normalized with the partition function goes towards Einstein $R^2$ scalar system with Lugendre transformation on $k[\mu\to 0^+(>\sqrt{z}/2),\Lambda]$.

\section{Functional renormalization group and non local heat kernel expansion}
\label{sec:2}
Functional renormalization group equation is the method for verify the couplings beta functions with the local heat kernel expansions \cite{Percacci}. The functional rfenormalization group equation is the equation with the cutoff function. The equation is found by the Wetterich \cite{W93}. Wetterich equation or called as the functional renormalization group equation is writte as,
\bea
\frac{d\Gamma_k}{dt}=\frac{1}{2}\tr\Big(\frac{\delta^2\Gamma_k}{\delta\phi\delta\phi}+R_k\Big)^{-1}\frac{dR_k}{dt}, \ \ \ t=\log k.
\eea
Where we denote the values and functions. $k$ is the cutoff momentum, $R_k$ is the cutoff function, the function $\phi(x)$ is the scalar function on the theoretical space which means the field expression of the theory at stand. This functional renormalization group equation is carry up the field with the integration of the cutoff momentum. We here consider the laplace operator with the factor $\bm U$,
\bea
\Delta=-D^2\bm 1+\bm U,
\eea
where $D_\mu$ is the covariant derivative in the space. To consider the expansion of the functional renormalization group equation, we begin with the action as,
\bea
S(\phi)=\frac{1}{2}\int d^dx\sqrt{g}\phi(x)(\Delta+\omega)\phi(x).
\eea
Be care for we thinking about the field theory as Eucledian space. The value $\omega$ is the constant. The citoff function $R_k$ is given by the eigen value of the laplace operator,
\bea
R_k=R_k(z\leftrightarrow \Delta).
\eea
$R_k(z)$ runs with $0<k<\Lambda \to \infty$. This function is cutoff function set by $k^2>z$ region. The cutoff action is written as,
\bea
\Delta S_k=\frac{1}{2}\int d^d x\sqrt{g}\phi(x)R_k(z)\phi(x).
\eea
Therefore total action is given as follows,
\bea
S(\phi)+\Delta S(\phi)=\frac{1}{2}\int d^d x\sqrt{g}\phi(x)(\Delta+\omega+R_k(\Delta))\phi(x).
\eea
We use the Eigen value $z$ at the needed situations. The weight $\sqrt{g}$ is apparent written here. The average effective aqction is given as,
\bea
\Gamma_k=\frac{1}{2}\log\Big[\det(\Delta+\omega+R_k(\Delta))\Big].
\eea
This derivation is done by the definition of the effective action from the field $\phi(x)$. The functional renormalization group equation is give as,
\bea
\frac{d\Gamma_k}{dt}=\frac{1}{2}\tr \frac{\pa_kR_k(\Delta)}{\Delta+\omega+R_k(\Delta)}.
\label{dgam}
\eea
The inverse propagator with the cutoff function is,
\bea
h_k(\Delta,\omega)=\frac{\pa_k R_k(\Delta)}{\Delta+\omega+R_k(\Delta)}.
\label{hk}
\eea
Here we introduce the laplace parameter $z$. The \p{dgam} with \p{hk} is the laplace transformed by,
\bea
\frac{d\Gamma_k}{dt}=\frac{1}{2}\int_0^\infty ds \til h_k(s)\tr e^{-s\Delta}.
\eea
$z$ is the real value from the sight of the eigenvalue. From the expression in \cite{CPRT16}, the heat kernel is expressed by,
\bea
\tr e^{-s\Delta}=\frac{1}{(4\pi s)^{d/2}}\int d^d x\sqrt{g}\tr\Big[1-s\bm U+s\bm 1\frac{R}{6}\nn
+s^2[\bm 1 R_{\mu\nu}f_{Ric}(s\Delta)R^{\mu\nu}+\bm 1 Rf_R (s\Delta)R
\nn
+Rf_{RU}(s\Delta)\bm U+\bm U f_U(s\Delta)\bm U+\Omega_{\mu\nu} f_\Omega(s\Delta)\Omega^{\mu\nu}]+...].
\eea
Where $\Omega_{\mu\nu}$ is the connection of the gravity coupled field on theoretical space. The space is defined the gravutational manifold $S$ and matter bandle $V$ as $S\times V$ connected with $\Omega_{\mu\nu}$. Structures written as $f_{\cdot}(x)$ is given as,
\bea
f(x)=\int_0^1 d\xi e^{-x\xi(1-\xi)},
\eea
and,
\bea
f_{Ric}(x)&=&\frac{1}{6x}+\frac{f(x)-1}{x^2},\nn
f_{R}(x)&=&\frac{f(x)}{32}+\frac{f(x)}{8x}-\frac{7}{48x}-\frac{f(x)-1}{8x^2},\nn
f_{RU}(x)&=&-\frac{f(x)}{4}-\frac{f(x)-1}{2x},\nn
f_U(x)&=&\frac{f(x)}{2},\nn
f_\Omega(x)&=&-\frac{f(x)-1}{2x}.
\eea
When we consider Taylor series with $x<<1$, we could expand $f(x)$ as,
\bea
f(x)=1-\frac{x}{6}+\frac{x^2}{60}+\mathcal O(x^4).
\eea
$f_{\cdot}(x)$ is computed easily. Here we use the eigen value expression $z$ and so write down the functional renormalization griup equation as follows,
\bea
\frac{d\Gamma_k}{dt}=\frac{1}{2(4\pi)^{d/2}}\int d^d x\sqrt{g}\tr\Big(\bm 1\Big[\int_0^\infty ds\til h_{k,\omega}(s)s^{-d/2}\Big]-\bm U\Big[\int_0^\infty ds\til h_{k,\omega}(s)s^{-d/2+1}\Big]
\nn
+\bm 1\frac{R}{6}\Big[\int_0^\infty ds\til h_{k,\omega}(s)s^{-d/2+1}\Big]+\bm 1R_{\mu\nu}\Big[\int_0^\infty ds \til h_{k,\omega}(s)s^{-d/2+2}f_{Ric}(sz)\Big]R^{\mu\nu}
\nn
+\bm 1 R\Big[\int_0^\infty ds \til h_{k,\omega}(s)s^{-d/2+2}f_R (sz)\Big]R+R\Big[\int_0^\infty ds\til h_{k,\omega}(s)s^{-d/2+2}f_{RU}(sz)\Big]\bm U
\nn
+\bm U\Big[\int_0^\infty ds \til h_{k,\omega}(s)s^{-d/2+2}f_U (sz)\Big]\bm U+\Omega_{\mu\nu}\Big[\int_0^\infty ds\til h_{k,\omega}(s)s^{-d/2+2}f_\Omega (sz)\Big]\Omega^{\mu\nu}+...\Big)
\eea
Here we introduce $Q_n[f]$ functionals with positive integars $n>0$. $Q_n[f]$ is given by,
\bea
Q_n[f]=\int_0^\infty ds s^{-n}\til f(s)=\frac{1}{\Gamma(n)}\int_0^\infty dw w^{n-1}f(w).
\eea
The last equal is Mellin transformation. We denote that $n=0$ is,
\bea
Q_0[f]=f(0).
\eea
In later we understand the $Q_n[f]$ includes resolutions $n=0,1,2,...$. Here we back to the definition of $Q_n$ function with $n=0,1,2,3,...$ as parallel transformation $a$. The functional is given by,
\bea
Q_n[h_{k,\omega}^a]=\int_0^\infty ds \til h_{k,\omega}(s)s^{-n}e^{-sa}, \ \ \ n=0,1,2,3,...
\eea
We denote the function $h_{k,\omega}^a$,
\bea
h_{k,\omega}^a (z)=h_{k,\omega}(z+a).
\eea
The convenient function is introduced as,
\bea
g_A=\int_0^\infty ds \til h_{k,\omega}(s)s^{-d/2+2}f_A(sz), \ \ \ n=0,1,2,3,...
\eea
$A$ index is meaning that $Ric,R,RU,U,\Omega$. Therefore, the functional renormalization group equation is non local heat kernel expanded as follows,
\bea
\frac{d\Gamma_k}{dt}=\frac{1}{2}\frac{1}{(4\pi)^{d/2}}\int d^d x\sqrt{g}\Big[Q_{d/2}[h_k]\tr\bm 1+Q_{d/2-1}[h_k]\tr\Big(\frac{R}{6}\bm 1-\bm U\Big)+R_{\mu\nu}g_{Ric}R^{\mu\nu}\tr\bm 1\nn
+Rg_R R\tr\bm 1+Rg_{RU}\tr\bm U+\tr(\bm U g_U \bm U)+\tr(\Omega_{\mu\nu}g_\Omega \Omega^{\mu\nu})+...\Big]
\eea
We wrote $h_k(z)=h_{k,\omega}(z)$. The $g_A$ functions are,
\bea
g_{Ric}&=&\frac{1}{6z}Q_{d/2-1}[h_k]-\frac{1}{z^2}Q_{d/2}[h_k]+\frac{1}{z^2}\int_0^1 d\xi Q_{d/2}[h_k^{z\xi(1-\xi)}],\nn
g_R&=&-\frac{7}{48z}Q_{d/2-1}[h_k]+\frac{1}{8z^2}Q_{d/2}[h_k]+\frac{1}{32}\int_0^1 d\xi Q_{d/2-2}[h_k^{z\xi(1-\xi)}]\nn
&+&\frac{1}{8z}\int_0^1 d\xi Q_{d/2-1}[h_k^{z\xi(1-\xi)}]-\frac{1}{8z^2}\int_0^1 d\xi Q_{d/2}[h_k^{z\xi(1-\xi)}],\nn
g_{RU}&=&\frac{1}{2z}Q_{d/2-1}[h_k]-\frac{1}{4}\int_0^1 d\xi Q_{d/2-2}[h_k^{z\xi(1-\xi)}]-\frac{1}{2z}\int_0^1 d\xi Q_{d/2-1}[h_k^{z\xi(1-\xi)}],\nn
g_U&=&\frac{1}{2}\int_0^1 d\xi Q_{d/2-2}[h_k^{z\xi(1-\xi)}],\nn
g_{\Omega}&=&\frac{1}{2z}Q_{d/2-1}[h_k]-\frac{1}{2z}\int_0^1 d\xi Q_{d/2-1}[h_k^{z\xi(1-\xi)}].
\eea
Remaining our work is computing $g_A$ functions with a set of the cutoff functions. To ensure this aim, we use the optimized cutoff function,
\bea
R_k(z)=(k^2-z)\theta(k^2-z)=k^2(1-\til z)\theta(1-\til z).
\eea
Where $\til z=z/k^2$ is the dimensionless property. When we consider $n>0$, we could use Mellin transformation. Therefore, $Q_n[h_k]$ is computed as,
\bea
Q_n[h_k]=\frac{1}{\Gamma(n)}\int_0^\infty dz z^{n-1}\frac{2k^2\theta(k^2-z)}{z+\omega+(k^2-z)}=\frac{2k^{2n}}{\Gamma(n+1)(1+\til\omega)}, \ \ \ n>0,
\eea
$n=0$ solution is included the last naturally,
\bea
Q_n[h_k]=\frac{2k^{2n}}{\Gamma(n+1)(1+\til\omega)}, \ \ \ n=0,1,2,...
\eea
Similarlly, we compute $Q_n[h_k^{z\xi(1-\xi)}]$. Thinking about $n>0$, the functional is,
\bea
Q_n[h_k^{z\xi(1-\xi)}]&=&\frac{1}{\Gamma(n)}\int_0^\infty dx x^{n-1}\frac{2k^2}{k^2+\omega}\theta(k^2-x-z\xi(1-\xi))\nn
&=&\frac{1}{\Gamma(n)}\int_0^{k^2-z\xi(1-\xi)}dxx^{n-1}\frac{2k^2}{k^2+\omega}\theta(k^2-z\xi(1-\xi))\nn
&=&\frac{1}{\Gamma(n+1)}\frac{2k^2}{k^2+\omega}(k^2-z\xi(1-\xi))^n\theta(k^2-z\xi(1-\xi)).
\eea
We know that the $n=0$ is included in above. The $Q_n[h_k^{z\xi(1-\xi)}]$ with $n=0,1,2,3,...$ is given by,
\bea
Q_n[h_k^{z\xi(1-\xi)}]=\frac{1}{\Gamma(n+1)}\frac{2k^2}{k^2+\omega}(k^2-z\xi(1-\xi))^n\theta(k^2-z\xi(1-\xi)), \ \ \ n=0,1,2,3,...
\label{fqx}
\eea
\p{fqx} is integrated from $\xi=0$ to $\xi=1$. To do the integration of \p{fqx}, we cosider the integration contour with $\xi-f(\xi)$ coordinate. Cross points are given by,
\bea
f(\xi)=k^2-z\xi(1-\xi)=0,
\eea
we find points $\alpha,\beta$ as,
\bea
\alpha=\frac{1}{2}\Big[1-\sqrt{1-\frac{4k^2}{z}}\theta(z-4k^2)\Big], \ \ \ 
\beta=\frac{1}{2}\Big[1+\sqrt{1-\frac{4k^2}{z}}\theta(z-4k^2)\Big].
\eea
The integration is carried with $k^2-z\xi(1-\xi)>0$. This means that the integration region is $0\leq \xi\leq \alpha$ and $\beta\leq \xi \leq 1$. From these reasons, we find $Q_n[h_k^{z\xi(1-\xi)}]$ integration as follows,
\bea
\int_0^1 d\xi Q_0[h_k^{z\xi(1-\xi)}]&=&\frac{2k^2}{k^2+\omega}\Big[1-\sqrt{1-\frac{4k^2}{z}}\theta(z-4k^2)\Big],\nn
\int_0^1 d\xi Q_1[h_k^{z\xi(1-\xi)}]&=&\frac{2k^4}{k^2+\omega}\Big[1-\frac{z}{6k^2}+\frac{z}{6k^2}\Big(1-\frac{4k^2}{z}\Big)^{3/2}\theta(z-4k^2)\Big],\nn
\int_0^1 d\xi Q_2[h_k^{z\xi(1-\xi)}]&=&\frac{k^6}{k^2+\omega}\Big[1-\frac{z}{3k^2}+\frac{z^2}{30k^4}-\frac{z^2}{30k^4}\Big(1-\frac{4k^2}{z}\Big)^{5/2}\theta(z-4k^2)\Big].
\eea
$g_A$ functions are calculated straightforwardlly,
\bea
g_{Ric}&=&\frac{1}{30}\frac{1}{1+\til\omega}\Big[1-\Big(1-\frac{4}{\til z}\Big)^{5/2}\theta(\til z-4)\Big],\nn
g_R&=&\frac{1}{1+\til\omega}\Big[\frac{1}{60}-\frac{1}{16}\Big(1-\frac{4}{\til z}\Big)^{1/2}\theta(\til z-4)\nn
&+&\frac{1}{24}\Big(1-\frac{4}{\til z}\Big)^{3/2}\theta(\til z-4)+\frac{1}{240}\Big(1-\frac{4}{\til z}\Big)^{5/2}\theta(\til z-4)\Big],\nn
g_{RU}&=&\frac{1}{1+\til\omega}\Big[-\frac{1}{3}+\frac{1}{2}\Big(1-\frac{4}{\til z}\Big)^{1/2}\theta(\til z-4)-\frac{1}{6}\Big(1-\frac{4}{\til z}\Big)^{3/2}\theta(\til z-4)\Big],\nn
g_U&=&\frac{1}{1+\til\omega}\Big[1-\Big(1-\frac{4}{\til z}\Big)^{1/2}\theta(\til z-4)\Big],\nn
g_\Omega&=&\frac{1}{6}\frac{1}{1+\til\omega}\Big[1-\Big(1-\frac{4}{\til z}\Big)^{3/2}\theta(\til z-4)\Big].
\eea
We derived master equation of the functional renormalization group equation with the non local heat kernel expansion. The master equation is convenient to verify the equation with the fields setups. The results are anticipated to have the coupled matters fields in the manifold quantum gravity. The sophisticated adjoints of the heat kernel coefficients and structures in stands after discussions.

\section{Gravity coupled with the scalar and appatrent normalized effective action}
\label{sec:3}
In this section we compute the effective action of the gravity coupled with scalar field theory in \cite{CPRT16}. After that, we normalized this effective action towards normalized effective action that would be matched with the appropriate tests confirmations in fine signs. The gravity coupled with the scalar theory is expanded as fluctuation perturbative expansions used in \cite{Percacci,CPRT16}. And so, we cosider the fields theory contributions with background metric and scalar field are fixed. The functional renormalization laplace operators are found by the fields setups include the ghost action. Heat kernel coefficients and structures are calculated to results, by the way, the ghost coefficients amplitudes are adjusted as adjoints. The effective action in $k^2>z/4=-\nabla^2/4$ to the ultraviolet area with $d=4$ in \cite{CPRT16}. Finally, we could have the normalization partition function makes the effective action runs from $\mu^2\to 0^{+2}>z/4$ to $\Lambda^2$ normalized into the Einstein $R^2$ action includes the coupled interaction from Lugendre transformation. We recognized that Einstein $R^2$ interacted action ensures the gravitational physics as the interacted gravity, parts are given in the interacted terms.

\subsection{Running effective action of the gravity and scalar field}
We begin with the preparations of the sctions. Einstein action is written by,
\bea
S_H[g+h]=-\frac{1}{\kappa}\int d^d x\sqrt{g+h}R(g+h).
\eea
$\kappa$ is defined as,
\bea
\kappa=16\pi G_0.
\eea
The scalar action is given by,
\bea
S_m[g+h,\phi]=\int d^d x\sqrt{g+h}\Big[\frac{1}{2}(g+h)^{\mu\nu}\nabla_\mu \phi\nabla_\nu \phi+V(\phi)\Big].
\eea
When we consider the fields theory, we should be gauge fixed with the appropriate gauge function. We use the gauge function called as De-Donder gauge function,
\bea
\chi_\mu=\nabla^\nu h_{\mu\nu}-\frac{1}{2}\nabla_\mu h.
\eea
The gauge fixing is expressed as the action,
\bea
S_{gf}[h,g]=\frac{1}{2}\int d^d x\sqrt{g}\chi_\mu \chi^\mu.
\eea
The gauge fixing is done with the gauge fixed action above. The gauge invariance is breaking with the selection of the gauge function. To recover the invariance, we introduce the possibility of the ghost action written by,
\bea
S_{gh}[\bar C,C,g]=\int d^d x\sqrt{g}\bar C_\mu(-\nabla^2 \delta^\mu{}_{\nu}-R^\mu{}_\nu)C^\nu.
\eea
The ghost laplace operator is,
\bea
(\Delta_{gh})^\mu{}_\nu=-\nabla^2 \delta^\mu{}_\nu-R^\mu{}_\nu.
\eea
Here we consider perturbative expansion of the scalar coupled gravity. Now we consider the perturbative expansion of the gravity as,
\bea
S_H(g+h)+S_{gf}(g+h)=S_H(g)+\frac{1}{2\kappa}\int d^d x\sqrt{g}h_{\alpha\beta}\mathcal H^{\alpha\beta\mu\nu}h_{\mu\nu}.
\eea
The gravitational hessian is given by,
\bea
\mathcal H^{\mu\nu}{}_{\rho\sigma}=K^{\mu\nu}{}_{\rho\sigma}(-\nabla^2)+u^{\mu\nu}{}_{\rho\sigma}.
\eea
De-Witt metric $K^{\mu\nu}{}_{\rho\sigma}$ and $u^{\mu\nu}{}_{\rho\sigma}$ are given by,
\bea
K^{\alpha\beta\mu\nu}=\frac{1}{2}\Big(\delta^{\alpha\beta,\mu\nu}-\frac{1}{2}g^{\alpha\beta}g^{\mu\nu}\Big)=\frac{1}{4}(g^{\mu\alpha}g^{\nu\beta}+g^{\nu\alpha}g^{\mu\beta}-g^{\alpha\beta}g^{\mu\nu}),
\eea
and so on,
\bea
u^{\mu\nu}{}_{\rho\sigma}=K^{\mu\nu}{}_{\rho\sigma}R+\frac{1}{2}(R^{\mu\nu}g_{\rho\sigma}+R_{\rho\sigma}g^{\mu\nu})-\delta^{(\mu}{}_{(\rho}R^{\nu)}{}_{\sigma)}-R^{(\mu}{}_{(\rho}{}^{\nu)}{}_{\sigma)}.
\eea
The matter action is expressed as,
\bea
S_m(g+h,\phi+\delta\phi)=\int d^d x\sqrt{g+h}\Big[\frac{1}{2}(g^{\mu\nu}-h^{\mu\nu}+h^{\mu}{}_{\rho}h^{\rho\nu})\nabla_\mu(\phi+\delta\phi)\nabla_\nu(\phi+\delta\phi)\nn
+V(\phi)+V'(\phi)\delta\phi+\frac{1}{2}V''(\phi)\delta\phi^2\Big].
\eea
The quadratic fields terme are remained because of the functional renormalization group equation detected by the twice any fields derivatives. Therefore, we should consider the contributions of the perturbative expansions as the quadratic fields. As we consider the perturbative expansions of the scalar coupled gravity, the recommanded form is the ,
\bea
\frac{1}{2\kappa}\int d^d x\sqrt{g}\delta\Phi^T \bm H\delta\Phi,
\eea
with field definition,
\bea
\delta \Phi=
\begin{pmatrix}
 h_{\mu\nu} \\ \delta\phi 
\end{pmatrix}
.
\eea
Fields are used as the vector notation and The hessian $\bm H$ is treated as the matrix as,
\bea
\bm H=
\begin{pmatrix}
  H^{\alpha\beta\mu\nu} & H^{\alpha\beta\cdot} \\
  H^{\cdot\mu\nu} & H^{\cdot\cdot}
\end{pmatrix}
.
\eea
The hessian is written as,
\bea
\bm H=\bm K(-\nabla^2)+2\bm V^\delta \nabla_\delta+\bm U.
\eea
Matrics $\bm K,\bm V^\delta, \bm U$ are computed towards,
\bea
\bm K=
\begin{pmatrix}
K^{\alpha\beta\mu\nu} & 0 \\
0 & \kappa
\end{pmatrix}
,
\eea
continually,
\bea
\bm V^\delta=
\begin{pmatrix}
0 & -\kappa K^{\alpha\beta\gamma\delta}\nabla_\gamma\phi \\
\kappa K^{\mu\nu\gamma\delta}\nabla_\gamma\phi & 0
\end{pmatrix}
,\\
\bm U=
\begin{pmatrix}
U^{\alpha\beta\mu\nu} & \frac{1}{2}\kappa g^{\alpha\beta}V'(\phi) \\
2\kappa K^{\mu\nu\gamma\delta}\nabla_\gamma \nabla_\delta \phi+\frac{1}{2}\kappa g^{\mu\nu}V'(\phi) & \kappa V''(\phi)
\end{pmatrix}
.
\eea
Where, $K^{\alpha\beta\mu\nu}$ is the De-Witt metric. The $U^{\alpha\beta\mu\nu}$ is expanded as follows,
\bea
U^{\alpha\beta\mu\nu}=K^{\alpha\beta\mu\nu}R+\frac{1}{2}(g^{\mu\nu}R^{\alpha\beta}+R^{\mu\nu}g^{\alpha\beta})-\frac{1}{4}(g^{\alpha\mu}R^{\beta\nu}+g^{\alpha\nu}R^{\beta\mu}+g^{\beta\mu}R^{\alpha\nu}+g^{\beta\nu}R^{\alpha\mu})\nn
-\frac{1}{2}(R^{\alpha\mu\beta\nu}+R^{\alpha\nu\beta\mu})+\kappa\Big[-\frac{1}{2}K^{\alpha\beta\mu\nu}(\nabla\phi)^2 -\frac{1}{4}(g^{\alpha\beta}\nabla^\mu \phi \nabla^\nu \phi+g^{\mu\nu}\nabla^\alpha \phi \nabla^\beta \phi)\nn
+\frac{1}{4}(g^{\alpha\mu}\nabla^\beta\phi\nabla^\nu\phi+g^{\alpha\nu}\nabla^\beta\phi\nabla^\mu\phi+g^{\beta\mu}\nabla^\alpha\phi\nabla^\nu\phi+g^{\beta\nu}\nabla^\alpha\phi\nabla^\mu\phi)-K^{\alpha\beta\mu\nu}V\Big].
\eea
From this hessian, we find the laplace operator as new relation. $\bm K^{-1}$ is prepared as,
\bea
\bm K^{-1}=
\begin{pmatrix}
K^{-1}_{\alpha\beta\mu\nu} & 0 \\
0 & \frac{1}{\kappa}
\end{pmatrix}
,
\eea
where we write the inverse De-Witt metric,
\bea
K^{-1}_{\alpha\beta\mu\nu}=2\delta_{\alpha\beta,\mu\nu}-\frac{2}{d-2}g_{\alpha\beta}g_{\mu\nu}.
\eea
The laplace operator is defined as $\bm K^{-1}\bm H$ as,
\bea
\Delta=\bm I(-\nabla^2)+2\bm Y^\delta \nabla_\delta+\bm W.
\eea
Also $\bm I$ is the unit matrix, matrixs $\bm Y^\delta$ and $\bm W$ are calculated in following results. $\bm Y^\delta$ is written as,
\bea
\bm Y^\delta=
\begin{pmatrix}
0 & -\kappa \delta_{\alpha\beta}{}^{\gamma\delta}\nabla_\gamma \phi \\
K^{\mu\nu\gamma\delta}\nabla_\gamma \phi & 0
\end{pmatrix}
.
\eea
$\bm W$ is computed towards,
\bea
\bm W=
\begin{pmatrix}
W_{\alpha\beta}{}^{\mu\nu} & -\frac{2}{d-2}\kappa g_{\alpha\beta}V'(\phi) \\
2K^{\mu\nu\gamma\delta}\nabla_\gamma \nabla_\delta\phi+\frac{1}{2}g^{\mu\nu}V'(\phi) & V''(\phi)
\end{pmatrix}
,
\eea
where we write $W_{\alpha\beta}{}^{\mu\nu}$,
\bea
W_{\alpha\beta}{}^{\mu\nu}=2U_{\alpha\beta}{}^{\mu\nu}-\frac{d-4}{d-2}g_{\alpha\beta}\Big[R^{\mu\nu}-\frac{1}{2}Rg^{\mu\nu}-\frac{\kappa}{2}\Big(\nabla^\mu \phi\nabla^\nu
\phi-\frac{1}{2}g^{\mu\nu}(\nabla\phi)^2\Big)\Big]-\frac{\kappa}{2}g_{\alpha\beta}g^{\mu\nu}V.~~~
\eea
This laplace operator is the different form from the definition. Then, we redefine the covariant derivative as,
\bea
\til\nabla_\mu=\nabla_\mu\bm I-\bm Y_\mu.
\eea
The laplace operator is expressed as,
\bea
\Delta=-\til\nabla_\mu\til\nabla^\mu+\til{\bm W}, \ \ \ \til{\bm W}=\bm W-\nabla_\mu\bm Y^\mu+\bm Y_\mu \bm Y^\mu.
\eea
To compute the $\til{\bm W}$, we prepare some parts of the laplace operator. $\nabla_\mu\bm Y^\mu$ is,
\bea
\nabla_\mu \bm Y^\mu=
\begin{pmatrix}
0 & -\kappa\nabla_\alpha \nabla_\beta\phi \\
K^{\mu\nu\gamma\delta}\nabla_\delta\nabla_\gamma \phi & 0
\end{pmatrix}
.
\eea
$\bm Y_\mu\bm Y^\mu$ is the following result,
\bea
\bm Y_\mu\bm Y^\mu=
\begin{pmatrix}
-\kappa \delta_{\alpha\beta}{}^{\gamma\delta}K^{\mu\nu\epsilon}{}_{\delta}\nabla_\gamma \phi\nabla_\epsilon \phi & 0 \\
0 & -\kappa(\nabla\phi)^2
\end{pmatrix}
.
\eea
We reach the $\til{\bm W}$ into,
\bea
\til{\bm W}=
\begin{pmatrix}
W_{\alpha\beta}{}^{\mu\nu}-\kappa \delta_{\alpha\beta}{}^{\gamma\delta}K^{\mu\nu\epsilon}{}_{\delta}\nabla_\gamma\phi\nabla_\epsilon\phi & -\frac{2}{d-2}\kappa g_{\alpha\beta}V'+\kappa\nabla_\alpha\nabla_\beta\phi \\
K^{\mu\nu\gamma\delta}\nabla_\gamma\nabla_\delta\phi+\frac{1}{2}g^{\mu\nu}V' & V''-\kappa(\nabla\phi)^2
\end{pmatrix}
.
\eea
Components of $\til{\bm W}$ are given by,
\bea
A_{\alpha\beta}{}^{\mu\nu}&=&W_{\alpha\beta}{}^{\mu\nu}-\frac{1}{2}\kappa \delta^{(\mu}{}_{(\alpha}\nabla_{\beta)}\nabla^{\nu)}\phi+\frac{1}{4}\kappa g^{\mu\nu}\nabla_\alpha\phi\nabla_\beta\phi,\nn
B_{\alpha\beta}&=&-\frac{2}{d-2}\kappa g_{\alpha\beta}V'+\kappa\nabla_\alpha \nabla_\beta\phi,\nn 
C^{\mu\nu}&=&K^{\mu\nu\alpha\beta}\nabla_\alpha \nabla_\beta\phi+\frac{1}{2}g^{\mu\nu}V',\nn
D&=&-\kappa(\nabla\phi)^2+V''.
\eea
From here, we consider the curvature expression. The curvature is calculted to,
\bea
\til \Omega_{\mu\nu}&=&[\nabla_\mu-\bm Y_\mu,\nabla_\nu-\bm Y_\nu]\nn
&=&\Omega_{\mu\nu}-\nabla_\mu\bm Y_\nu+\nabla_\nu\bm Y_\mu+[\bm Y_\mu,\bm Y_\nu].
\eea
Some parts of terms above are verified towards,
\bea
\nabla_{[\rho}\bm Y_{\sigma]}=
\begin{pmatrix}
0 & -\kappa\delta_{\alpha\beta}{}^{\gamma\delta}g_{[\sigma|\delta}\nabla_{\rho]}\nabla_\gamma\phi \\
K^{\mu\nu\gamma\delta}g_{[\sigma|\delta}\nabla_{\rho]}\nabla_\gamma\phi & 0
\end{pmatrix}
,\\
\bm Y_{[\rho}\bm Y_{\sigma]}=
\begin{pmatrix}
-\kappa\delta_{\alpha\beta,\gamma[\rho}K^{\mu\nu}{}_{|\epsilon|\sigma]}\nabla^\gamma\phi\nabla^\epsilon\phi & 0 \\
0 & 0
\end{pmatrix}
.
\eea
Components of the curvature matrix are to come results as follows,
\bea
(\til \Omega_{\rho\sigma})_{\alpha\beta}{}^{\mu\nu}&=&(\Omega_{\rho\sigma})_{\alpha\beta}{}^{\mu\nu}-2\kappa\delta_{\alpha\beta,\gamma[\rho}K^{\mu\nu}{}_{|\epsilon|\sigma]}\nabla^\gamma\phi\nabla^\epsilon\phi,\nn
(\til \Omega_{\rho\sigma})_{\alpha\beta}&=&2\kappa\delta_{\alpha\beta}{}^{\gamma\delta}g_{[\sigma|\delta}\nabla_{\rho]}\nabla_\gamma\phi,\nn
(\til \Omega_{\rho\sigma})^{\mu\nu}&=&-2K^{\mu\nu\gamma\delta}g_{[\sigma|\delta}\nabla_{\rho]}\nabla_\gamma\phi.
\eea
We prepared the laplace operator properties and curvature properties already. Begin to expand the functional renormalization group equation with the non local heat kernel expressions.

We consider heat kernel coefficients and non local structure terms with calculetion results of the needed properties. $d=4$ is applied in latter here. Remember the laplace operator of scalar coupling gravity means that $\til{\bm W}$ is the unique tensor. Then, we compute some needs. The trace is,
\bea
\tr\til{\bm W}=A_{\mu\nu}{}^{\mu\nu}+D,
\eea 
where,
\bea
A_{\mu\nu}{}^{\mu\nu}=6R-\kappa(\nabla\phi)^2-10\kappa V.
\eea
$\tr\til{\bm W}^2$ is calculated as,
\bea
\tr\til{\bm W}^2=A_{\mu\nu}{}^{\alpha\beta}A_{\alpha\beta}{}^{\mu\nu}+2B_{\alpha\beta}C^{\alpha\beta}+D^2,
\eea
where,
\bea
A_{\mu\nu}{}^{\alpha\beta}A_{\alpha\beta}{}^{\mu\nu}&=&3R_{\mu\nu\rho\sigma}R^{\mu\nu\rho\sigma}-6R_{\mu\nu}R^{\mu\nu}+5R^2-\frac{3}{2}\kappa R(\nabla\phi)^2-12\kappa VR\nn
&+&\frac{7}{4}\kappa^2((\nabla\phi)^2)^2+2\kappa^2 V(\nabla\phi)^2+10\kappa^2 V^2,\\
B_{\alpha\beta}C^{\alpha\beta}&=&\kappa V' \nabla^2\phi-2\kappa V'^2+\frac{1}{2}\kappa \nabla_\mu\nabla_\nu\phi\nabla^\mu \nabla^\nu\phi-\frac{1}{4}\kappa(\nabla^2\phi)^2.
\eea
Continually, curvature notations are computed as follows,
\bea
\tr\til \Omega_{\mu\nu}\til \Omega^{\mu\nu}=(\til\Omega_{\mu\nu})_{\alpha\beta}{}^{\gamma\delta}(\til \Omega^{\mu\nu})_{\gamma\delta}{}^{\alpha\beta}+2(\til \Omega_{\mu\nu})_{\alpha\beta}(\til \Omega^{\mu\nu})^{\alpha\beta},
\eea
where,
\bea
(\til \Omega_{\mu\nu})_{\alpha\beta}{}^{\gamma\delta}(\til \Omega^{\mu\nu})_{\gamma\delta}{}^{\alpha\beta}&=&-6R_{\mu\nu\rho\sigma}R^{\mu\nu\rho\sigma}+\kappa R(\nabla\phi)^2+2\kappa R_{\mu\nu}\nabla^\mu \phi\nabla^\nu \phi-\frac{3}{2}\kappa^2((\nabla\phi)^2)^2,~~~~~\\
2(\til\Omega_{\mu\nu})_{\alpha\beta}(\til\Omega^{\mu\nu})^{\alpha\beta}&=&\kappa(\nabla^2\phi)^2-4\kappa\nabla_\mu\nabla_\nu\phi\nabla^\mu\nabla^\nu\phi.
\eea
To simplify the heatkernel coefficients and structures, we use below relation,
\bea
R_{\mu\nu\rho\sigma}R^{\mu\nu\rho\sigma}=4R_{\mu\nu}R^{\mu\nu}-R^2.
\eea
Heat kernel coefficients of the scalar coupling gravity are list as,
\bea
\tr \bm 1=11, \ \ \ \tr\bm 1\frac{R}{6}-\tr\til{\bm W}=-\frac{25}{6}R+2\kappa(\nabla\phi)^2-V''+10\kappa V.
\label{gshk}
\eea
The ghost heat kernel coefficients are computed as,
\bea
\tr \bm 1_{gh}=4, \ \ \ \tr\bm 1_{gh}\frac{R}{6}-\tr\bm U_{gh}=\frac{5}{3}RC_{gh,2}=\frac{1}{3}R.
\label{ghhk}
\eea
The trace is adjusted with the amplitude constant $C_{gh,2}=\frac{1}{5}$. Remaining our work is calculating non local structures. The non local structures of the gravity and scalar are,
\bea
R_{\mu\nu}(11f_{Ric}+6f_U-24f_\Omega)R^{\mu\nu}+R(11f_R+6f_{RU}+2f_U+6f_\Omega)R\nn
+\kappa R(\nabla\phi)^2\Big(-2f_{RU}-\frac{3}{2}f_U+f_\Omega\Big)+2\kappa R^{\mu\nu}\nabla_\mu\phi\nabla_\nu\phi f_\Omega+RV'' f_{RU}+\kappa RV(-10f_{RU}-12f_U)\nn
+\kappa\nabla_\mu\nabla_\nu\phi\nabla^\mu\nabla^\nu\phi(f_U-4f_\Omega)+\kappa(\nabla^2\phi)^2\Big(-\frac{1}{2}f_U+f_\Omega\Big)+2\kappa\nabla^2\phi V' f_U-2\kappa V''(\nabla\phi)^2f_U\nn
+2\kappa^2V (\nabla\phi)^2f_U-4\kappa V' V' f_U+V''V''f_U+\kappa^2((\nabla\phi)^2)^2\Big(\frac{11}{4}f_U-\frac{3}{2}f_\Omega\Big)+10\kappa^2VVf_U.~~~~~
\label{gss}
\eea
The non local structures of the ghost are,
\bea
R_{\mu\nu}(4f_{Ric}+f_U-4f_\Omega)R^{\mu\nu}+R(4f_R-f_{RU}+f_\Omega)R.
\label{ghs}
\eea
From these results, we know used heat kernel coefficients and non local structure as expansions of the functional renormalization group. Therefore, we expand the functional renormalization group equation below with related properties,
\bea
\frac{d\Gamma_k}{dt}=\frac{1}{2}\tr\frac{\pa_t R_k(\Delta)}{\Delta+R_k(\Delta)}-\tr\frac{\pa_t R_k(\Delta_{gh})}{\Delta_{gh}+R_k(\Delta_{gh})}.
\eea
Using \p{gshk}, \p{ghhk}, \p{gss} and \p{ghs}, we fond the functional renormalization group equation as follows,
\bea
\frac{d\Gamma_k}{dt}=\frac{1}{32\pi^2}\int d^ 4 x\sqrt{g}\Big[3k^4+k^2\Big(-\frac{29}{3}R+4\kappa(\nabla\phi)^2-2V''+20\kappa V\Big)+R_{\mu\nu}g_1R^{\mu\nu}+Rg_2 R \nn
+\kappa R(\nabla\phi)^2 g_3+\kappa R_{\mu\nu}\nabla^\mu\phi \nabla^\nu\phi g_4+RV''g_5+\kappa RV g_6+\kappa\nabla_\mu\nabla_\nu\phi\nabla^\mu\nabla^\nu\phi g_7\nn
+\kappa((\nabla^2\phi))^2 g_8+\kappa^2((\nabla\phi)^2)^2 g_9+\kappa\nabla^2\phi V'g_{10}-\kappa(\nabla\phi)^2V'' g_{10}+\kappa^2(\nabla\phi)^2 Vg_{10}\nn
-2\kappa V' V' g_{10}+\frac{1}{2}V'' V'' g_{10}+5\kappa^2 VVg_{10}\Big].~~~~~
\eea
Where we defined $g_a, a=1,2,3,4,5,6,7,8,9,10$ functions. These structures are given by,
\bea
g_1&=&3f_{Ric}+4g_U-16g_\Omega,\\
g_2&=&3g_R +2g_U +8g_{RU}+4g_\Omega,\\
g_3&=&-2g_{RU}-\frac{3}{2}g_U+g_\Omega,\\
g_4&=&2g_\Omega,\\
g_5&=&g_{RU},\\
g_6&=&-10g_{RU}-12g_U,\\
g_7&=&g_U-4g_\Omega,\\
g_8&=&-\frac{1}{2}g_U+g_\Omega,\\
g_9&=&\frac{11}{4}g_U-\frac{3}{2}g_\Omega,\\
g_{10}&=&2g_U.
\eea
Here we consider to derive the effective action with results of structures $g_a, a=1,2,3,4,5,6,7,8,9,10$. To ensure our purposses, $g$-structures are calculated with $k^2>z/4$,
\bea
g_1=\frac{43}{30}, \ \ g_2=\frac{1}{20}, \ \ g_3=-\frac{2}{3}, \ \ g_4=\frac{1}{3}, \ \ g_5=-\frac{1}{3},\\
g_6=-\frac{26}{3}, \ \ g_7=\frac{1}{3}, \ \ g_8=-\frac{1}{3}, \ \ g_9=\frac{5}{2}, \ \ g_{10}=2.
\eea
We are assuming that the $z>0$ and $k^2>z/4$ runnings. We set the integration region $k[\mu\to 0^+(>\sqrt z/2),\Lambda]$ as the ultraviolet cutoff momentum $\Lambda$, and for, we could write down the effective action with needed sources,
\bea
\Gamma_\Lambda[g]=\frac{1}{32\pi^2}\int d^4 x\sqrt{g}\Big[3\Lambda^4+\Big(-\frac{29}{3}R+4\kappa(\nabla\phi)^2-2V''+20\kappa V\Big)\Lambda^2\nn
+\Big(\frac{43}{30}R_{\mu\nu}R^{\mu\nu}+\frac{1}{20}R^2-\frac{2}{3}\kappa R(\nabla\phi)^2-\frac{26}{3}\kappa RV-\frac{1}{3}RV''\nn
+\frac{5}{2}\kappa^2((\nabla\phi)^2)^2+2\kappa^2V(\nabla\phi)^2-4\kappa V''(\nabla\phi)^2\nn
+10\kappa^2 V^2-4\kappa V'^2+V''^2\Big)\log\Big(\frac{\Lambda^2}{\mu^2}\Big)+\textrm{finites}\Big].
\label{ea}
\eea
Be care for the renormalization scale $\mu^2$ runs to $0^{+2}>z/4$. The full effective action is written by some apparent descriptions from the finite terms.

\subsection{Normalization of the effective action}
The effective action including renormalization scale and cutoff. We are normalize this effective action \p{ea} with the definition of quantum corrected action being equal to the effective action normalized by the cutoff path integration normalization as,
\bea
Z=\int_{\mu\to 0^+}^\Lambda (dk)N[g,\phi]\exp(-S_k[g,\phi]), \ \ \ \mu^2\to 0^{+2}>z/4.
\eea
The path integral is the summed ups of the cutoff scale running from $\mu\to \Lambda$ as the definition of the total partition function of $Z$. The quantum corrected action is the effective action \p{ea} as,
\bea
S_k[g,\phi]=\frac{1}{32\pi^2}\int d^4 x\sqrt{g}\Big[3k^4+\Big(-\frac{29}{3}R+4\kappa(\nabla\phi)^2-2V''+20\kappa V\Big)k^2\nn
+\Big(\frac{43}{30}R_{\mu\nu}R^{\mu\nu}+\frac{1}{20}R^2-\frac{2}{3}\kappa R(\nabla\phi)^2-\frac{26}{3}\kappa RV-\frac{1}{3}RV''\nn
+\frac{5}{2}\kappa^2((\nabla\phi)^2)^2+2\kappa^2V(\nabla\phi)^2-4\kappa V''(\nabla\phi)^2\nn
+10\kappa^2 V^2-4\kappa V'^2+V''^2\Big)\log\Big(\frac{k^2}{\mu^2\to 0^{+2}(>z/4)}\Big)+\textrm{finites}\Big].
\label{qca}
\eea
The normalize factor $N[g,\phi]$ is the important property to make $Z$ towards the basic fixed exponential part. To do so, we consider the factor as,
\bea
N[g,\phi]=\frac{N_0[g,\phi]}{N_{v}[g,\phi]}.
\eea
$N_0[g,\phi]$ is the basic exponential as to include the bare action having Einstein scalar coupling action. $N_v[g,\phi]$ is take $\int_{\mu\to 0^+(>\sqrt z/2)}^\Lambda (dk)\exp(-S_k[g,\phi])$ towards $1$. It isn't permitted to make $N_0[g,\phi]$ arbitrally. $N_0[g,\phi]$ should includes every relating action terms with thinking about the well defined action in gravity. The quantum corrected action is true in \p{qca}, and on, we could derive the $N_0[g,\phi]$ as follows,
\bea
N_0[g,\phi]=\exp\Bigg[-\frac{1}{32\pi^2}\int d^4 x\sqrt{g}\Big[(const.)+\Big(-\frac{29}{3}RC_1+4\kappa(\nabla\phi)^2C_2-2V''C_3+20\kappa VC_4\Big)\nn
+\Big(\frac{43}{30}R_{\mu\nu}R^{\mu\nu}C_a+\frac{1}{20}R^2C_b-\frac{2}{3}\kappa R(\nabla\phi)^2C_c-\frac{26}{3}\kappa RVC_d-\frac{1}{3}RV''C_e\nn
+\frac{5}{2}\kappa^2((\nabla\phi)^2)^2C_f+2\kappa^2V(\nabla\phi)^2C_g-4\kappa V''(\nabla\phi)^2C_h\nn
+10\kappa^2 V^2C_i-4\kappa V'^2C_j+V''^2C_k\Big)\Big]\Bigg].~~~~~
\eea
Introduced coefficients are $C_1,C_2,C_3,C_4$ dimensionfull parameters and $C_a,...,C_k$ dimensionless parameters. For true Einstein scalar theory, we know that the parameters as,
\bea
C_n>0 \ \ and \ \ n=1,2,3,4,\\
C_m=(-1)^{T,F}C_M(>0) \ \ and \ \ m=M=a,b,c,...,k.
\eea
Therefore, the normal factor $N_0[g,\phi]$ is reexpressed by,
\bea
N_0[g,\phi]=\exp\Bigg[-\frac{1}{32\pi^2}\int d^4 x\sqrt{g}\Big[(const.)+\Big(-\frac{29}{3}RC_1+4\kappa(\nabla\phi)^2C_2-2V''C_3+20\kappa VC_4\Big)\nn
+\Big(\frac{43}{30}R_{\mu\nu}R^{\mu\nu}(-1)^{T,F}C_A+\frac{1}{20}R^2(-1)^{T,F}C_B-\frac{2}{3}\kappa R(\nabla\phi)^2(-1)^{T,F}C_C-\frac{26}{3}\kappa RV(-1)^{T,F}C_D\nn
-\frac{1}{3}RV''(-1)^{T,F}C_E+\frac{5}{2}\kappa^2((\nabla\phi)^2)^2(-1)^{T,F}C_F+2\kappa^2V(\nabla\phi)^2(-1)^{T,F}C_G\nn
-4\kappa V''(\nabla\phi)^2(-1)^{T,F}C_H+10\kappa^2 V^2(-1)^{T,F}C_I-4\kappa V'^2(-1)^{T,F}C_J+V''^2(-1)^{T,F}C_K\Big)\Big]\Bigg].~~~~~
\eea
The partition function called by connected is,
\bea
W=\log Z=-\frac{1}{32\pi^2}\int d^4 x\sqrt{g}\Big[(const.)+\Big(-\frac{29}{3}RC_1+4\kappa(\nabla\phi)^2C_2-2V''C_3+20\kappa VC_4\Big)\nn
+\Big(\frac{43}{30}R_{\mu\nu}R^{\mu\nu}(-1)^{T,F}C_A+\frac{1}{20}R^2(-1)^{T,F}C_B-\frac{2}{3}\kappa R(\nabla\phi)^2(-1)^{T,F}C_C-\frac{26}{3}\kappa RV(-1)^{T,F}C_D\nn
-\frac{1}{3}RV''(-1)^{T,F}C_E+\frac{5}{2}\kappa^2((\nabla\phi)^2)^2(-1)^{T,F}C_F+2\kappa^2V(\nabla\phi)^2(-1)^{T,F}C_G\nn
-4\kappa V''(\nabla\phi)^2(-1)^{T,F}C_H+10\kappa^2 V^2(-1)^{T,F}C_I-4\kappa V'^2(-1)^{T,F}C_J+V''^2(-1)^{T,F}C_K\Big)\Big].~~~
\eea
From these results, we could find the effective gravitational action as follows,
\bea
\Gamma[g,\phi]=\frac{1}{32\pi^2}\int d^4 x\sqrt{g}\Big[(const.)+\Big(-\frac{29}{3}RC_1+4\kappa(\nabla\phi)^2C_2-2V''C_3+20\kappa VC_4\Big)\nn
+\Big(\frac{43}{30}R_{\mu\nu}R^{\mu\nu}(-1)^{T,F}C_A+\frac{1}{20}R^2(-1)^{T,F}C_B-\frac{2}{3}\kappa R(\nabla\phi)^2(-1)^{T,F}C_C-\frac{26}{3}\kappa RV(-1)^{T,F}C_D\nn
-\frac{1}{3}RV''(-1)^{T,F}C_E+\frac{5}{2}\kappa^2((\nabla\phi)^2)^2(-1)^{T,F}C_F+2\kappa^2V(\nabla\phi)^2(-1)^{T,F}C_G\nn
-4\kappa V''(\nabla\phi)^2(-1)^{T,F}C_H+10\kappa^2 V^2(-1)^{T,F}C_I-4\kappa V'^2(-1)^{T,F}C_J+V''^2(-1)^{T,F}C_K\Big)\Big].~~~
\label{gsef}
\eea
The effective gravitational action has many important terms of the gravity, scalar field theory and coupled fields. The renormalization is to induced in the $N_v[g,\phi]$ confirmlly. From the considerationsof the gravity, the gravitational effective action is used as the $R^2$ gravitation with non scalar field as,
\bea
\Gamma[g,0]=\frac{1}{32\pi^2}\int d^4 x\sqrt{g}\Big((const.)-\frac{29}{3}RC_1
+\frac{43}{30}R_{\mu\nu}R^{\mu\nu}(-1)^{T,F}C_A+\frac{1}{20}R^2(-1)^{T,F}C_B\Big).~~~~~
\label{gref}
\eea
This is the Einstein $R^2$ pure action for consider the gravitation developments in the universal story. The scalar coupling Einstein $R^2$ action is anticipated in considering the early gravitational ages in physics.

\section{Results and future}
We derived the gravitational effective action from the quantum corrected action with running the cuoff $\mu(\to 0^+>\sqrt{z}/2)\to\Lambda$. The actions \p{gsef} and \p{gref} were basic actions for gravitation since the Einstein $R^2$ couplling with scalar matter and potential. Therefore, we could lead the useful results of the gravity from the quantum gravity what is the non local heat kernel expansions in the functional renormalization group equation. 

In section 2, we prepared to verify the functional renormalization group equation with the quantum gravity. We considered the scalar action on the curved space. The functional renormalization group equation was expressed by the propagator also expanded to the non local heat kernel expansions these are written by heat kernel coefficients and structure functions.  As the result, the nonlocal heat kernel expanssion of the functional renormalization group equation was lead towards the expanded results given by $Q_n$ functionals. $Q_n$ functionals were calculated to local result and non local structure dependent result. Then, the master equation was structed formally in this section.

In latter section, we considered that the renormalization group are expanded with the gravity coupled with the scalar matter on Einstein manifold. The perturbative expanssions were giving to the scalar coupled gravity hessian, on the other side, the ghost operator were also given. The functional renormalization group was expanded by these laplace operators towards the quantum corrected Wetterich equation. The heat kernel coefficients and structure functions were calculated with this quantum corrected equation, therefore, we reached the effective action included the cutoff momentum $\Lambda$ with the scale $\mu^2>z/4$. Since we got the quantum corrected action with the cuoff $k$, we could normalize it towards non cutoff scale dependent effective aqction as \p{gsef} and \p{gref}. Therefore, we knew that the gravitational effective action lead from normalization of the quantum corrected action in the quantum gravity to ensure the Einstein $R^2$ scalar coupled or non coupled systems.


\end{document}